\def\pmb#1{\setbox0=\hbox{#1}%
\kern-.025em\copy0\kern-\wd0	
\kern-.05em\copy0\kern-\wd0
\kern-.025em\raise.0433em\box0}

\def \bomega {{\pmb{$\omega$}}}

\documentclass{elsart}
\usepackage{graphicx}

\usepackage{amssymb}

\begin{document}



\begin {frontmatter}
\title{Polarized Foreground Emission from Dust:
Grain Alignment and MHD Turbulence}
\author[Alex]{A. Lazarian \& J. Cho},
\address[Alex]{University of Wisconsin-Madison,
Astronomy Department, 475 N. Charter St., Madison, WI 53706, e-mail:
lazarian@astro.wisc.edu}



\begin{abstract}
Aligned grains present a foreground for cosmic microwave emission
studies. We review basic physical
processes involved in grain alignment and discuss the niches for
different alignment mechanisms. We show that
mechanisms which were favored for
decades do not look so promising right now, while the
radiative torque mechanism ignored for more than 20 years looks 
quite attractive. We discuss alignment of small rotating
grains that are thought to be responsible for 10-90~GHz 
anomalous foreground and polarization arising from magnetic grains.

\end{abstract}
\end{frontmatter}

\section{Introduction}

Diffuse Galactic microwave emission carries important information on 
the fundamental
properties of interstellar medium, but it also interferes with the
Cosmic Microwave Background (CMB) experiments 
(see Tegmark et al. 1999 and references
therein). Polarization of the CMB provides information about the Universe
that is not contained in the temperature data (see Prunet \& Lazarian
1999, Davis \& Wilkinson 1999) and a number of groups
around the world (see Table~1 in Staggs et al. 1999) work hard to determine
the CMB polarization. In view of this work, the issue of determining the
degree and properties of Galactic foreground polarization becomes vital.

A known source of polarized foreground is thermal emission from
aligned grains.
Grain alignment of interstellar dust 
has been discovered more than half a century ago. Hall (1949) and 
Hiltner (1949) reported polarization that 
was attributed to the differential extinction of starlight by
dust particles with longer axes preferentially aligned.
Very soon it was realized that the alignment happens with
respect to the interstellar magnetic field\footnote{The relation between
grain alignment direction and that of magnetic field is clear from
a comparison of synchrotron polarization maps and those of galactic starlight
polarization (see Serkowski, Mathewson \& Ford 1975). Recent measurements
of polarization in external galaxies (see Jones 2000) makes this relation
even more obvious.}.

For many years grain alignment theory had a very limited predictive power
and was an issue of hot debates. 
This caused somewhat cynical approach to the theory among some of the
polarimetry practitioners who preferred to be guided in their work
by the following rules of thumb: {\it All grains are always aligned 
and the alignment happens with the longer grain axes perpendicular to
magnetic field.} This simple recipe was shattered, however, by observational
data which indicated that \\
I. Grains of sizes smaller than the critical size are either not aligned
or marginally aligned (Mathis 1986, Kim \& Martin 1995).\\
II. Carbonaceous grains are not aligned, but silicate grains are aligned
(see Mathis 1986).\\  
III. Substantial part of 
grains deep within molecular clouds are not aligned (Goodman et al. 1995,
Lazarian, Goodman \& Myers 1997).\\
VI. Grains might be aligned with longer axes parallel to 
magnetic fields\footnote{A simple, but not always clearly understood
property of grain alignment in interstellar medium is that it always 
happens  in respect to magnetic field. It can be shown that the
fast (compared with other time scales) Larmor
precession of grains makes the magnetic field the reference axis. 
Note, however,
that grains
 may align with their longer axes {\it perpendicular} or {\it parallel}
to magnetic field direction.
Similarly, magnetic fields may change their configuration and orientation
in space (e.g. due to Alfven waves), but if the time for such
a change is much longer than the Larmor period the alignment of
grains {\it in respect to the field lines} persists as the consequence
of preservation of the adiabatic invariant.} (Rao et al 1998).

A  boost of the interest to grain alignment  came from the
search of Cosmic Microwave Background (CMB) polarization (see
Lazarian \& Prunet 2002, for a review). Aligned dust in this case acts as
a source of a ubiquitous 
foreground that is necessary to remove from the data. It is clear
that understanding of grain alignment is the key element for such a removal.
With the present level of interest to the CMB polarization 
we are bound to have a lot of microwave and far infrared polarimetry
data. It is important to understand to what extend this data reflects
the structure of magnetic field in the Galaxy and whether this
data can be used to get insight into the processes of galactic magnetic
field generation and into interstellar turbulence\footnote{Velocity
and magnetic field statistics provide the most clear insight in
what is going on with the turbulence. With velocity statistics available
through the recently developed Velocity Channel Analysis (VCA) technique (
Lazarian \& Pogosyan 2000) magnetic fields statistics is the missing
element. Polarized starlight and emission from aligned grains
provide the easiest way to get such a statistics.}. 

Alignment of the tiny spinning grains that are thought to be responsible
for the microwave emission in the 10-90~GHz is a separate topic that
we deal with. We also discuss polarization arising from grains with
strong magnetic response and an interpretation of the polarized emission
in terms of galactic MHD turbulence.

\section{Basic Facts: Aligned Grains \& Polarized Radiation}

\subsection{Linear Polarized Starlight from Aligned Grains}

For an ensemble of aligned grains the extinction perpendicular 
and parallel to
the direction
of alignment and parallel are different\footnote{According
to Hildebrand \& Dragovan (1995) the best fit of the grain properties
corresponds to oblate grains with the ratio of axis about 2/3.}. Therefore
 that is initially unpolarized starlight acquires polarization while
passing through a volume with aligned grains.
If the extinction in the direction of alignment is $\tau_{\|}$ and in
the perpendicular direction is  $\tau_{\bot}$
one can write the polarization, $P_{abs}$, by selective extinction
 of grains
as 
\begin{equation}
P_{abs}=\frac{e^{-\tau_{\|}}-e^{-\tau_{\bot}}}{e^{-\tau_{\|}}+e^{-\tau_{\bot}}}
\approx -{(\tau_{\|}-\tau_{\bot})}/2~,
\label{Pabs}
\end{equation}
where the latter approximation is valid for $\tau_{\|}-\tau_{\bot}\ll 1$.
To relate the difference of extinctions to the properties of aligned grains
one can take into 
account the fact that the extinction is proportional to the product
of the grain density and  their cross sections. If a cloud is composed of 
identical aligned grains
$\tau_{\|}$ and $\tau_{\bot}$ are proportional to the number of grains
along the light path times the corresponding cross sections, which
are, respectively, 
$C_{\|}$ and $C_{\bot}$.

In reality one has to consider additional complications like
incomplete grain alignment, and variations in the direction
of the alignment axis (in most cases the latter is the direction of
 magnetic field, as discussed above) along the line of sight.  
 To obtain 
an adequate description one can (see Roberge \& Lazarian 1999) consider
 an electromagnetic wave propagating along the line of sight
{\mbox{$\hat{\bf z}^{\bf\rm o}$}} axis.
The transfer equations for the Stokes parameters
depend on the cross sections,  $C_{xo}$ and $C_{yo}$, for linearly polarized
waves with the electric vector,  {\mbox{\boldmath$E$}}, 
along the {\mbox{$\hat{\bf x}^{\bf\rm o}$}} and 
{\mbox{$\hat{\bf y}^{\bf\rm o}$}} directions
that are in the plane perpendicular to {\mbox{$\hat{\bf z}^{\bf\rm o}$}}
(see Lee \& Draine 1985).

To calculate  $C_{xo}$ and $C_{yo}$,
one transforms the components of {\mbox{\boldmath$E$}} to
a frame aligned with the principal axes of the grain and
takes the appropriately-weighted sum of the
cross sections, $C_{\|}$ and , $C_{\bot}$ for {\mbox{\boldmath$E$}}
 polarized along the grain
axes (Fig~1b illustrates the geometry of observations).
When the transformation is carried out and the resulting
expressions are averaged over precession angles, one finds (see
transformations in Lee \& Draine 1985 for spheroidal grains and in
Efroimsky 2002 for a general case)
that
the mean cross sections are
\begin{equation}
C_{xo} = C_{avg} + \frac{1}{3}\,R\,\left(C_{\bot}-C_{\|}\right)\,
       \left(1-3\cos^2\zeta\right)~~~,
\label{eq-2_5}
\end{equation}
\begin{equation}
C_{yo} = C_{avg} + \frac{1}{3}\,R\,\left(C_{\bot}-C_{\|}\right)~~~,
\label{eq-2_6}
\end{equation}
where $\zeta$ is the angle between the polarization axis and the 
{\mbox{$\hat{\bf x}^{\bf\rm o}$}} {\mbox{$\hat{\bf y}^{\bf\rm o}$}}
plane;
$C_{avg}\equiv\left(2 C_{\bot}+ C_{\|}\right)/3$ is the effective
cross section for randomly-oriented grains.
To characterize the alignment we used in eq.~(\ref{eq-2_6})
 the Raylegh reduction factor
(Greenberg 1968)
\begin{equation}
R\equiv \langle G(\cos^2\theta) G(\cos^2\beta)\rangle
\label{R}
\end{equation}
where angular brackets denote ensemble averaging, $G(x) \equiv 3/2 (x-1/3)$,
 $\theta$ is the angle between the axis of the largest moment of inertia
(henceforth the axis of maximal inertia) and the magnetic field ${\bf B}$, while
$\beta$ is the angle between the angular momentum ${\bf J}$ and ${\bf B}$. 
 To characterize
${\bf J}$ alignment in grain axes and in respect to magnetic field,
 the
measures ${Q_X\equiv \langle G(\theta)\rangle}$ and 
$Q_J\equiv \langle G(\beta)\rangle$
are used.
Unfortunately, these statistics 
are not independent and therefore $R$ is not equal to $Q_J Q_X$ (see 
Lazarian 1998, Roberge
\& Lazarian 1999). This considerably complicates 
the treatment of grain alignment.

\subsection{Polarized Emission from Aligned Grains}

The difference in $\tau_{\|}$ and $\tau_{\bot}$ results in emission of
aligned grains being polarized:
\begin{equation}
P_{em}=\frac{(1-e^{-\tau_{\|}})-(1-e^{-\tau_{\bot}})}{(1-e^{-\tau_{\|}})+
(1-e^{-\tau_{\bot}})}\approx \frac{\tau_{\|}-\tau_{\bot}}
{\tau_{\|}+\tau_{\bot}}~,
\label{Pem}
\end{equation}
where both the optical depths $\tau{\|}$ are $\tau_{\bot}$ were
assumed to be small. Taking into account that 
both $P_{em}$ and $P_{abs}$ are functions
of wavelength $\lambda$ and 
combining eqs.(\ref{Pabs}) and (\ref{Pem}), one gets for
$\tau=(\tau_{\|}+\tau_{\bot})/2$
\begin{equation}
P_{em}(\lambda) \approx -P_{abs}(\lambda)/\tau(\lambda)~,
\label{Pem}
\end{equation}
which establishes the relation between polarization in emission and
absorption. The minus sign
in eq~(\ref{Pem})
reflects the fact that emission and absorption polarization are
orthogonal. 
As $P_{abs}$ depends on $R$, $P_{em}$ also depends on the Rayleigh reduction
factor.

\section{Grain Alignment Theory: New and Old Ideas}

We have seen in the previous sections that both linear and circular
polarizations depend on the degree of grain alignment given by
$R$-factor (\ref{R}). Therefore it is the goal of grain alignment theory
to determine this factor. The complexity of the grain alignment is
illustrated in Fig~1, which shows that grain alignment is indeed
a multi-stage process.

A number of different mechanisms that produce grain alignment has been
developed by now (see table~1 in Lazarian, Goodman \& Myers 1997). Dealing
with a particular situation one has to identify the dominant alignment process.
Therefore it is essential to understand different mechanisms. By now
the theory of grain alignment is rather complex. This makes it advantageous
to follow the evolution of grain alignment ideas. It is instructive
to see the major role that observations played in shaping up of the theory.

\begin{figure}[h]
\includegraphics[width=2.5in]{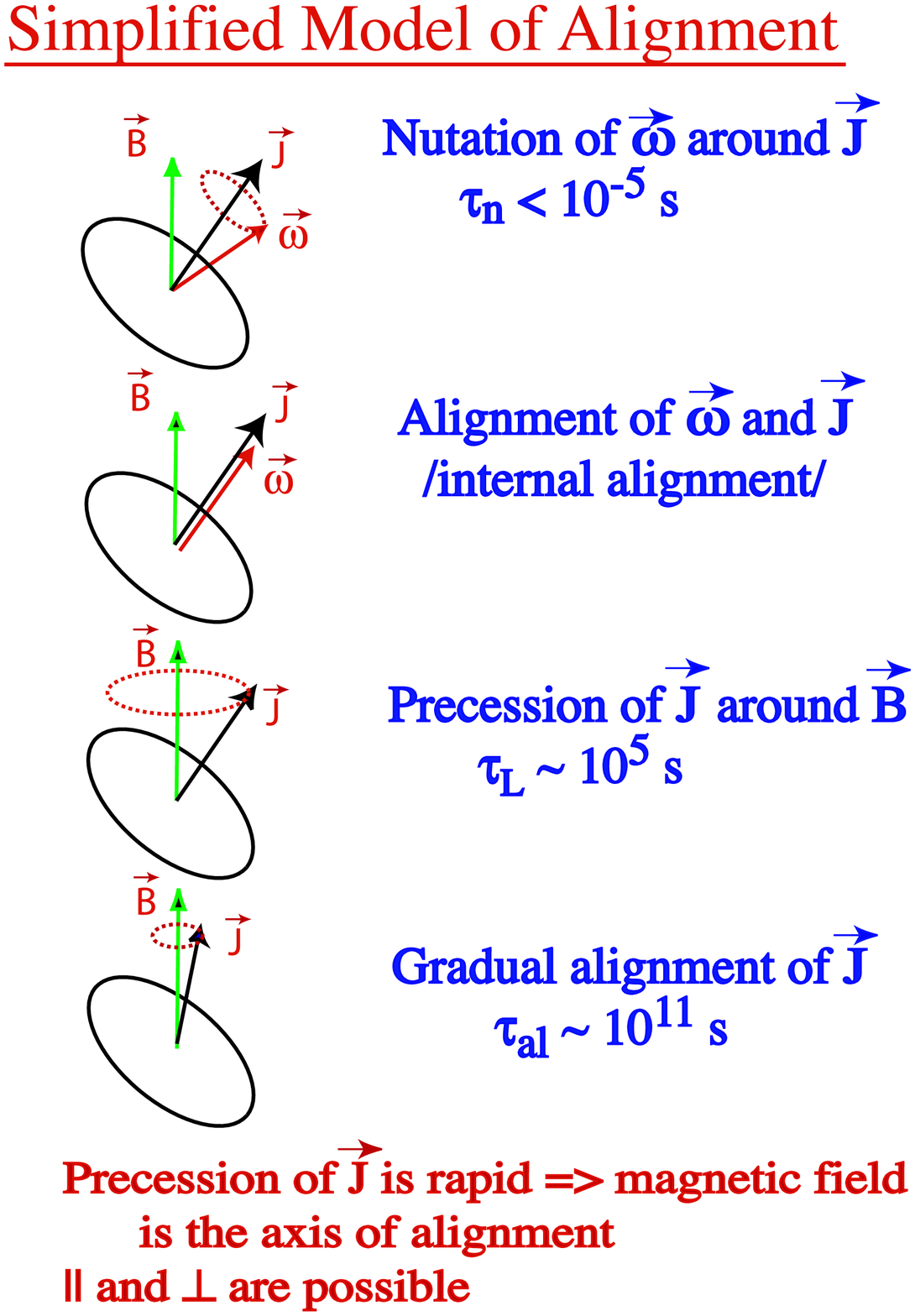}
\hfill
\includegraphics[width=3.in]{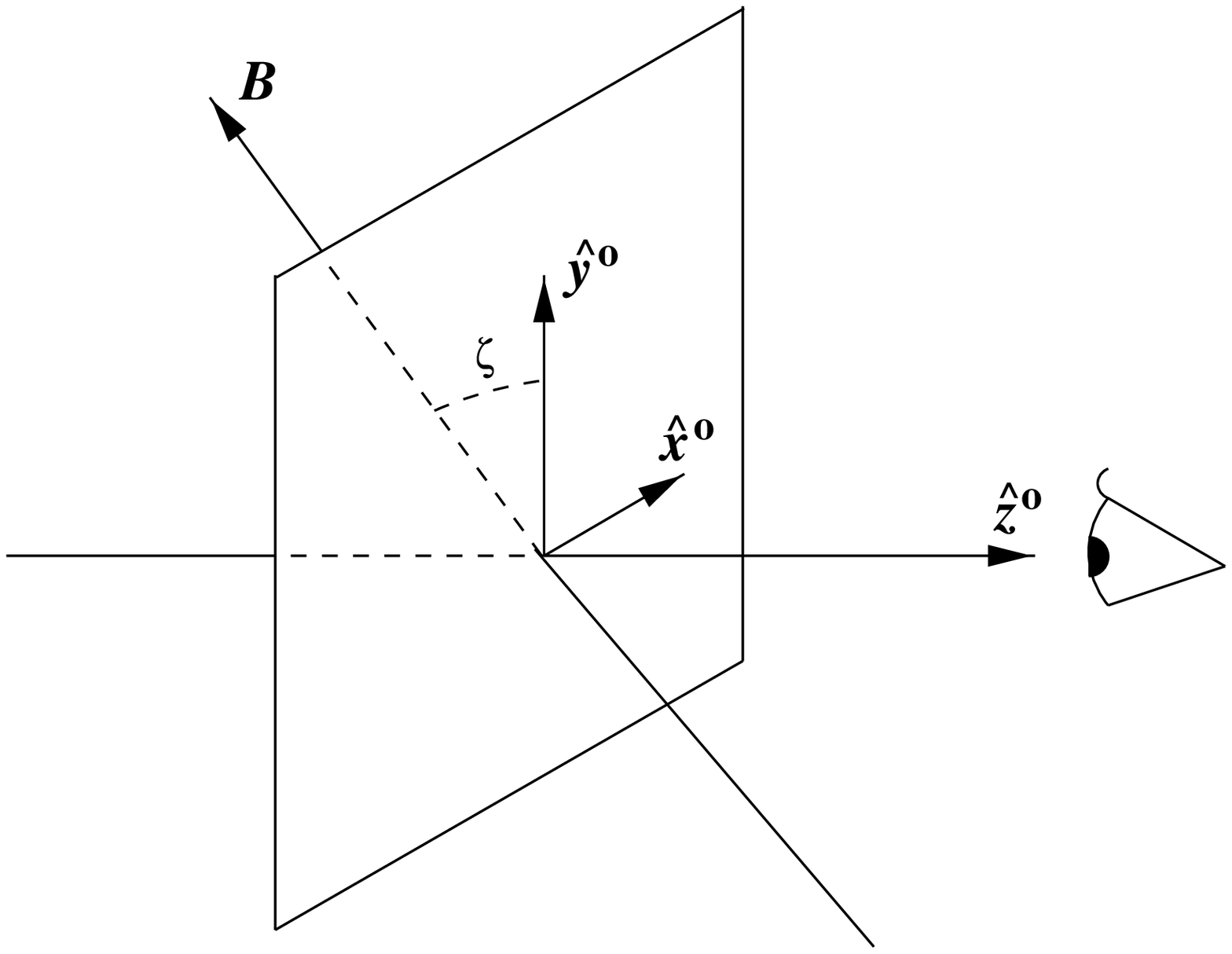}
\caption{a)Left panel. Alignment of grains implies several alignment
processes acting simultaneously spaning over many scales. 
Internal alignment was introduced by Purcell (1979) and was assumed to be
a slow process. Lazarian \& Draine (1999a) showed that the internal alignment
is $10^6$ times faster if nuclear spins are accounted for.  The time
scale of ${\bf J}$ and ${\bf B}$ alignment is given for diffuse interstellar
medium. It is faster in circumstellar regions and for comet dust.
 b) Right panel.
Geometry of observations (after Roberge \& Lazarian 1999).} 
\label{fig:2Dspek}
\end{figure}

\subsection{Foundations of the Theory}

The first stage of alignment theory development
started directly after the discovery of
starlight polarization. 
Nearly simultaneously Davis \& Greenstein (1950) 
and Gold (1951) proposed their scenarios of alignment. 

{\it Paramagnetic Alignment: Davis-Greenstein Process}\\
Davis-Greenstein
mechanism (henceforth D-G mechanism)
is based on the paramagnetic dissipation that is experienced
by a rotating grain. Paramagnetic materials contain unpaired
electrons which get oriented by the interstellar magnetic field ${\bf B}$. 
The orientation of spins causes
grain magnetization and the latter 
varies as the vector of magnetization rotates
 in grain body coordinates. This causes paramagnetic loses 
at the expense of grain rotation energy.
Note, that if the grain rotational velocity ${\bomega}$
is parallel to ${\bf B}$, the grain magnetization does not change with time
and therefore
no dissipation takes place. Thus the
paramagnetic dissipation  acts to decrease the component of ${\bomega}$
perpendicular to ${\bf B}$ and one may expect that eventually
grains will tend to rotate with ${\bomega}\| {\bf B}$
provided that the time of relaxation $t_{D-G}$ is much shorter than  $t_{gas}$,
the
time of randomization through chaotic gaseous bombardment.
In practice, the last condition is difficult to satisfy. For $10^{-5}$ cm 
grains
in the diffuse interstellar medium
$t_{D-G}$ is of the order of $7\times 10^{13}a_{(-5)}^2 B^{-2}_{(5)}$s , 
while  $t_{gas}$ is $3\times 10^{12}n_{(20)}T^{-1/2}_{(2)} a_{(-5)}$ s (
see table~2 in Lazarian \& Draine 1997) if
magnetic field is $5\times 10^{-6}$ G and
temperature and density of gas are $100$ K and $20$ cm$^{-3}$, respectively. 
However, in view of uncertainties in
interstellar parameters the D-G theory initially looked plausible.

{\it Mechanical Alignment: Gold Process}\\
Gold mechanism is a process of mechanical alignment of grains. Consider
a needle-like grain interacting with a stream of atoms. Assuming
that collisions are inelastic, it is easy to see that every
bombarding atom deposits angular momentum $\delta {\bf J}=
m_{atom} {\bf r}\times {\bf v}_{atom}$ with the grain, 
which is directed perpendicular to both the
needle axis ${\bf r}$ and the 
 velocity of atoms ${\bf v}_{atom}$. It is obvious
that the resulting
grain angular momenta will be in the plane perpendicular to the direction of
the stream. It is also easy to see that this type of alignment will
be efficient only if the flow is supersonic\footnote{Otherwise grains
will see atoms coming not from one direction, but from a wide cone of
directions (see Lazarian 1997a) and the efficiency of 
alignment will decrease.}.
Thus the main issue with the Gold mechanism is to provide supersonic
drift of gas and grains. Gold originally proposed collisions between
clouds as the means of enabling this drift, but later papers (Davis 1955) 
showed that the process could  only align grains over limited patches of
interstellar space, and thus the process
cannot account for the ubiquitous grain 
alignment in diffuse medium.

{\it Quantitative Treatment and Enhanced Magnetism}\\
The first detailed analytical treatment of the problem of D-G
alignment was given by Jones \& Spitzer (1967) who described the alignment
of ${\bf J}$
using a Fokker-Planck equation. This 
approach allowed them to account for magnetization fluctuations
within grain material and thus provided a more accurate picture of 
${\bf J}$ alignment.
$Q_X$ was assumed to follow
the Maxwellian distribution, although the authors noted
that this might not be correct. 
The first numerical treatment of
D-G alignment was presented by Purcell (1969). 
By that time it became clear that the D-G
mechanism is too weak to explain the observed grain alignment. However,
Jones \& Spitzer (1967) noticed that if interstellar grains
contain superparamagnetic, ferro- or ferrimagnetic (henceforth SFM) 
inclusions\footnote{The evidence for such inclusions was found much later
through the study of interstellar dust particles captured in
the atmosphere (Bradley 1994).}, the
$t_{D-G}$ may be reduced by orders of magnitude. Since $10\%$ of
atoms in interstellar dust are iron
the formation of magnetic clusters in grains was not far fetched
(see Spitzer \& Turkey 1950, Martin 1995)
and therefore the idea was widely accepted. Indeed, with enhanced 
magnetic susceptibility the D-G mechanism was able to solve
all the contemporary problems of alignment. The conclusive
at this stage was the paper by Purcell \& Spitzer (1971) where
all various models of grain alignment, including, for
instance, the model of cosmic ray alignment 
by Salpeter \& Wickramasinche (1969)
and photon alignment by Harwit (1970)  
were quantitatively discussed and the D-G model with enhanced
magnetism was endorsed. It is this stage of development that is widely
reflected in many textbooks.

\subsection{Additional Essential Physics}

{\it Barnett Effect and Fast Larmor Precession}\\
It was realized by Martin (1971) that rotating charged grains will develop
magnetic moment and the interaction of this moment with the interstellar
magnetic field will result in grain precession. The characteristic
time for the precession was found to be comparable with $t_{gas}$. 
However, soon  a process that
renders much larger magnetic moment was discovered (Dolginov \& Mytrophanov 
1976). This process is the 
Barnett effect, which is converse of the Einstein-de Haas effect.
If in Einstein-de Haas effect a paramagnetic body starts rotating
 during remagnetizations
as its flipping 
electrons transfer the angular momentum (associated with their spins)
 to the
lattice, in the Barnett effect
the rotating body shares its angular momentum with the electron
subsystem  causing magnetization. The magnetization
is directed along the grain angular velocity and the value
of the Barnett-induced magnetic moment is $\mu\approx 10^{-19}\omega_{(5)}$~erg
gauss$^{-1}$ (where $\omega_{(5)}\equiv \omega/10^5{\rm s}^{-1}$). Therefore
the Larmor precession has a period $t_{Lar}\approx 3\times 10^6 B_{(5)}^{-1}$~s and 
the magnetic field defines the axis of alignment as we explained in section~1.

{\it Suprathermal Paramagnetic Alignment: Purcell Mechanism}\\
The next step was done by Purcell(1975, 1979),
who discovered that grains can rotate much faster than were previously
thought. He noted 
that variations of photoelectric yield, the H$_2$ formation efficiency,
and  variations of accommodation coefficient over grain surface
would result in uncompensated torques acting upon a
grain. The H$_2$ formation on the grain surface clearly illustrates the
process we talk about: if H$_2$ formation takes place only over particular
catalytic sites, these sites act as miniature rocket engines
spinning up the grain. Under such uncompensated torques the grain will spin-up to
velocities much higher than thermal (Brownian) and Purcell termed those
velocities ``suprathermal''. Purcell also noticed that for suprathermally
rotating grains
internal relaxation will bring ${\bf J}$ parallel to the axis of maximal
inertia (i.e. $Q_X=1$). Indeed, for an oblate spheroidal
grain with 
angular momentum ${\bf J}$  the energy can be written
\begin{equation}
E(\theta)=\frac{J^2}{I_{max}}\left(1+\sin^2\theta (h-1)\right)
\label{e}
\end{equation}
where $h=I_{max}/I_{\bot}$ is the ratio of the maximal to minimal moments
of inertia. Internal forces cannot change the angular momentum, but
it is evident from Eq.~(\ref{e}) that the energy can be decreased by
aligning the axis of maximal inertia along ${\bf J}$, i.e. decreasing
$\theta$. Purcell (1979) discusses two possible causes of internal 
dissipation,
the first one related to the well known inelastic relaxation, the second is
due to the mechanism that he discovered and termed ``Barnett relaxation''.
This process may be easily understood. We know that a 
freely rotating grain preserves the direction of
${\bf J}$, while angular velocity precesses about 
${\bf J}$ and in grain body axes.
We learned earlier that the Barnett effect results in the magnetization
vector parallel to $\bomega$. As a result, the Barnett magnetization
will precess in body axes and cause paramagnetic relaxation.
The ``Barnett equivalent magnetic field'', i.e. the equivalent external
magnetic field that would cause the same magnetization of the grain  
material, is $H_{BE}=5.6 \times10^{-3} \omega_{(5)}$~G, 
which is much larger than the interstellar magnetic 
field. Therefore the Barnett relaxation happens on the scale $t_{Bar}\approx
4\times 10^7 \omega_{(5)}^{-2}$~sec,
i.e. essentially instantly compared to $t_{gas}$ and $t_{D-G}$.

{\it Theory of Crossovers}\\
If $Q_X=1$ and the suprathermally rotating grains are immune to randomization
by gaseous bombardment, will paramagnetic grains be perfectly aligned with
$R=1$? This question was addressed by Spitzer \& McGlynn (1979)
(henceforth
SM79) who observed
that adsorption of heavy elements on a grain  should result
in the resurfacing phenomenon that, e.g.  should remove early sites
of H$_2$ formation and create new ones. As the result,
H$_2$ torques will occasionally change their direction and spin the grain
down. SM79  showed that in the absence of
random torques the spinning down grain will
 flip over preserving the direction of its original angular momentum.
However, in the presence of random torques
 this direction will be altered with the maximal deviation inflicted
over a short period of time just before and after the flip, i.e.
during the time when the value of grain angular momentum is minimal.
The actual value of angular momentum during this critical
period depends on the ability of ${\bf J}$ to deviate from
the axis of maximal inertia.
SM79 observed that as the Barnett relaxation 
couples ${\bf J}$ with the axis of maximal inertia it 
makes randomization of grains during crossover nearly complete. With the
resurfacing time $t_{res}$ estimated by SM79 to be of the order of $t_{gas}$,
the gain of the alignment efficiency was
insufficient to reconcile the theory and observations unless the grains 
had SFM inclusions.

{\it Radiative Torques}\\
If the introduction of the concept of suprathermality by Purcell changed
the way researchers thought of grain dynamics, the introduction of radiative torques
passed essentially unnoticed. Dolginov (1972) argued that quartz grains
may be spun up due to their specific rotation of polarization
while later Dolginov \& Mytrophanov (1976)
discovered that irregular grain shape may allow grains scatter left and right
hand polarized light differentially,  thus spinning up helical grains through
scattering of photons\footnote{The principal difference between radiative
torque mechanism and the radiative emission/absorption mechanism proposed
by Harwit (1970) is that the radiative torques are regular and therefore
increase the grain velocity in proportion to time. Harwit's mechanism,
on the other hand, is based on stochastic spin-up and therefore is
subdominant. We also note that the emission of photons is insensitive
to grain helicity as the emitted photons have wavelengths much larger than
the grain size.}. They stressed that the most efficient spin-up
is expected when grains size is comparable with the wavelength and estimated
the torque efficiency for particular helical grain shapes, but failed
to provide estimates of the relative efficiency of the mechanism in the
standard interstellar conditions. In any case, this ingenious idea had not
been appreciated for another 20 years.

{\it Observational tests: Serkowski Law}\\
All in all, by the end of seventies the the following alignment mechanisms
were known:
1. paramagnetic( a. with SFM inclusions,
   b. with suprathermal rotation),
2. mechanical,
3. radiative torques.
The third was ignored, the second was believed to be suppressed
for suprathermally rotating grains, which left
the two modifications of the paramagnetic mechanism as competing alternatives.
Mathis (1986) noticed that the interstellar polarization-wavelength dependence
known as the Serkowski law (Serkowski et al. 1975) can be explained if
grains larger that $\sim 10^{-5}$~cm are aligned, while smaller grains
are not. To account for this behavior Mathis (1986) noticed
that the SFM inclusions will have a better chance to
be in larger rather than smaller grains. The success of fitting
observational data persuaded
the researchers that the problem of grain alignment is solved at last.

\subsection{Present Stage of Grain Alignment Theory}

Optical and near infrared observations by 
Goodman et al. (1995)  showed 
that polarization efficiency may
drop within dark clouds while far infrared observations by Hildebrand et al. (1984),
Hildebrand et al. (1990)
revealing aligned grains within star-forming dark clouds. This
renewed interest to grain alignment problem.

{\it New Life of Radiative Torques}\\
Probably the most dramatic change of the picture was the unexpected advent
of radiative torques. Before Bruce Draine realized that the torques
can be treated with the versatile discrete dipole approximation (DDA)
code ( Draine \& Flatau 1994), their role was unclear. For instance, earlier on
difficulties associated with the analytical approach to
the problem were discussed in Lazarian (1995a).
However, very soon after that Draine (1996) modified the DDA code
 to calculate the torques acting on grains of arbitrary
shape. His work revolutionized the field! 
The magnitude of torques were found to be substantial and present
for grains of various irregular shape (Draine 1996, Draine \& Weingartner
1996). After that it became impossible
to ignore these torques. Being related to grain shape, rather than surface
these torques are long-lived\footnote{In the case of the Purcell's
rockets the duration of torque action is limited by the time of resurfacing,
while in the case of radiative torques it is the time scale on which the
grain is either destroyed via collisions, coagulates with another grain
or gets a different shape in the process of growth.}, i.e. 
$t_{spin-up}\gg t_{gas}$, 
which allowed Draine \& Weingartner (1996)
to conclude that in the presence of isotropic radiation the radiative 
torques can support fast grain rotation long enough in order for
paramagnetic torques to align grains (and without any SFM
inclusions). However, the important question was what would happen
in the presence of anisotropic radiation. Indeed, in the presence
of such radiation the torques will change as the grain aligns
 and this may result in a spin-down. Moreover,
anisotropic flux of radiation will deposit angular momentum 
which is likely to overwhelm rather weak paramagnetic torques. These sort of
questions were addressed by Draine \& Weingartner (1997) and it was
found that for most of the tried grain shapes the torques tend to 
align ${\bf J}$ along magnetic field. The reason for that is yet unclear
and some caution is needed as the existing treatment ignores the dynamics
of crossovers which is  very important for the alignment of
suprathermally rotating grains. Nevertheless, radiative torques
are extremely appealing as their predictions are consistent
with observational data (see Lazarian, Goodman \& Myers 1994, Hildebrand et 
al. 1999, see section 4 as well).

{\it New Elements of Crossovers}\\
Another unexpected development was a substantial change of the picture
of crossovers. As we pointed out earlier,  Purcell's discovery of
fast internal dissipation resulted in a notion that ${\bf J}$
should always stay along the axis of maximal inertia as long
as $t_{dis}\ll t_{gas}$. Calculations in
SM79 were based on this notion.
However, this
perfect coupling
 was questioned in Lazarian (1994) (henceforth L94), where it was shown that
thermal fluctuations within grain material partially randomize
the distribution of grain axes in respect to ${\bf J}$. 
The process was quantified in Lazarian \& Roberge (1997)
(henceforth LR97),
where the distribution of $\theta$ for a freely
rotating grain was defined through the Boltzmann distribution
 $\exp(-E(\theta)/kT_{grain})$,
where the energy $E(\theta)$ is given by Eq.~(\ref{e}). This finding
changed the understanding of crossovers a lot. First of all,
Lazarian \& Draine (1997)(henceforth LD97) observed that  thermal
fluctuations partially decouple ${\bf J}$ and the axis of maximal
inertia and therefore the value of angular moment at the moment
of a flip is substantially larger than SM79 assumed. Thus the
randomization during a crossover is  reduced and  LD97 obtained
a nearly
perfect alignment  for interstellar grains
rotating suprathermally, provided that
the grains were larger than a certain critical size $a_c$.  The latter 
size was found by
equating the time of the crossover and the time of the internal
dissipation $t_{dis}$. For $a<a_c$
Lazarian \& Draine (1999a) found new physical effects, which they termed
``thermal flipping'' and ``thermal trapping''. The thermal flipping
 takes place
as the time of the crossover becomes larger than $t_{dis}$.           
In this situation thermal fluctuations will enable flipovers. However,
being random, thermal fluctuations are likely to produce not a single
flipover, but multiple ones. As the grain flips back and forth the
regular (e.g. H$_2$) torques average out and the
grain can spend a lot of time rotating with thermal velocity, i.e.
being ``thermally trapped''. The paramagnetic alignment of 
grains rotating with 
thermal velocities is small (see above) 
and therefore grains with $a<a_{c}$ are
expected to be marginally aligned. The picture of preferential
alignment of large grains, as we know, corresponds to the Serkowski
law and therefore the real issue is to find the value of $a_c$.
The Barnett relaxation\footnote{A study by Lazarian \& Efroimsky (1999)
corrected the earlier estimate by Purcell (1979), but left the conclusion
about the Barnett relaxation
dominance, and therefore the value of $a_c$, intact. For larger objects, e.g.
for astreroids, comets, the inelastic relation is dominant (Efroimsky 
\& Lazarian 2000, Efroimsky 2001).}
 provides a comforting value of $a_c\sim 10^{-5}$~cm. However, in a
recent paper Lazarian \& Draine (1999b) reported a new solid state effect
that they termed ``nuclear relaxation''. This is an analog of Barnett
relaxation effect that deals with nuclei. Similarly to unpaired electrons
nuclei tend to get oriented in a rotating body. However the nuclear analog
of ``Barnett equivalent'' magnetic field is much larger and Lazarian \&
Draine (1999b) concluded that the nuclear relaxation can be a million times
faster than the Barnett relaxation. If this is true $a_c$ becomes of the
order $10^{-4}$~cm, which means that the majority of interstellar grains
undergo constant flipping and rotate essentially thermally in spite of
the presence of
uncompensated Purcell torques. The radiative torques
are not fixed in body coordinates and it is likely that they can provide
a means for suprathermal rotation for grains that are larger than the
wavelength of the incoming radiation. Naturally, it is of utmost importance
to incorporate the theory of crossovers into the existing codes, 
and this work is under way.

{\it New Ideas and Quantitative Theories}\\
An interest to grain alignment resulted in search of new mechanisms. For
instance, Sorrell (1995a,b) proposed a mechanism of grain spin-up due to
interaction with cosmic rays that locally heat grains and provide evaporation
of adsorbed H$_2$ molecules. However, detailed
calculations in Lazarian \& Roberge (1997b)
showed that the efficiency of the torques was overestimated; the observations
(Chrysostomou et al. 1996) did not confirm Sorrell's predictions either. 
A more promising idea that
ambipolar diffusion can align interstellar grains was
put forward in Roberge \& Hanany (1990) (calculations are done
in Roberge et al. 1995). Within this mechanism ambipolar
drift provides the supersonic velocities necessary for mechanical alignment.   
Independently L94 proposed a mechanism of mechanical grain alignment
using Alfven waves. Unlike the ambipolar diffusion, this mechanism 
operates even in ideal MHD and relies only on the difference in inertia of
atoms and grains and on the direct interaction of grains with fluctuating
magnetic field (Lazarian \& Yan 2002, Yan \& Lazarian 2002). 
An additional boost to interest to mechanical
processes was gained when it was shown that suprathermally rotating
grains can be aligned mechanically (Lazarian 1995a; Lazarian \& Efroimsky 1996;
Lazarian, Efroimsky \& Ozik 1996; Efroimsky 2002).
As it was realized that thermally rotating grains do not ${\bf J}$ tightly
coupled with the axis of maximal inertia (L94) and the effect
was quantified (LR97), it got possible to formulate quantitative theories
of Gold (Lazarian 1997a) and Davis-Greenstein (Lazarian 1997b, Roberge \& 
Lazarian 1999) alignments. Together with a better understanding of
grain superparamagnetism (Draine \& Lazarian 1998a),
damping of grain rotation (Draine \& Lazarian 1998b) and resurfacing of
grains (Lazarian 1995c), these developments
increased the predictive power of the grain alignment theory.

\section{Polarization of Spinning Grain Emission}

All the studies above dealt with classical ``large'' grains. What  about
very small (e.g. $a<10^{-7}$~cm) grains? Can they be aligned? The answer
to this question became acute after Draine \& Lazarian (1998a,b) explained
the anomalous galactic emission in the range $10-100$~GHz as arising
from rapidly (but thermally!) 
spinning tiny grains. This rotational dipole emission will
be polarized if grains are aligned.

Microwave emission from spinning grains is expected to be polarized if
grains are aligned. Alignment of ultrasmall grains which are
essentially large molecules is likely to be different from alignment 
of large (i.e. $a>10^{-6}$~cm) grains for which the
theory of grain alignment (see review by Lazarian 2000) has been developed.

One of the mechanisms that might produce alignment of the ultrasmall
grains is the
paramagnetic dissipation mechanism suggested half a century ago
by Davis and Greenstein (1951) as a means of explaining the
polarization of starlight. The Davis-Greenstein alignment mechanism is
straightforward: for a spinning grain
the component of interstellar magnetic field
perpendicular to the grain angular velocity varies in grain coordinates,
resulting in time-dependent magnetization, associated
energy dissipation, and a torque acting on the grain.
As a result grains  tend to rotate
with angular momenta parallel to the 
interstellar magnetic field.

Are the ultrasmall grains paramagnetic? The answer to this question
is positive owing to the presence of free radicals, paramagnetic
carbon rings  (see Altshuler
\& Kozyrev 1964) and captured ions.
For paramagnetic grains, the alignment time-scale
 $\tau\approx
10^4$~yr~$ (a/10^{-6}~{\rm cm} )^2 (10^{-13}~{\rm s}/K)$ with
$K(\omega)\equiv {\rm Im}(\chi)/\omega$, where $\omega$ is the 
angular rotational
velocity and $\chi(\omega)$ is the magnetic susceptibility.
The characteristic time of grain magnetic response is
the electron precession time in the field of its neighbors, which
is also called spin-spin relaxation time and is denoted $\tau_2$.
If $\omega \leq \tau_2^{-1}\sim
10^8$~s$^{-1}$, normal materials at $T\approx 20$~K
have $K\approx 10^{-13}$~s. For higher frequencies , however, $K(\omega)$
begins to decrease rapidly (Draine \& Lazarian 1999, henceforth DL99). 
As discussed earlier spinning grains must rotate
much faster to account for the Foreground X, thus apparently calling into 
question the efficacy of alignment by paramagnetic dissipation.

Lazarian \& Draine (2000, henceforth LD00) found 
that the traditional picture
of paramagnetic relaxation is 
incomplete, since it
disregards the splitting of energy levels within a rotating
body. 
Unpaired
electrons spin parallel and antiparallel
to the grain angular velocity have different energies causing
 the so-called ``Barnett magnetization'' (Landau \& Lifshitz 1960).
The Barnett effect, the inverse of the Einstein-De Haas effect, 
consists of the spontaneous magnetization of
a paramagnetic body
rotating in field-free space. This effect can be understood in
terms of the lattice sharing part of its angular momentum with 
the spin system. Therefore the implicit assumption in Davis 
\& Greenstein (1951)-- 
that the magnetization within a {\it rotating grain} in a {\it static} 
magnetic field is equivalent to the magnetization within a 
{\it stationary grain} in a {\it rotating} magnetic field --
is clearly not exact.

If electrons within a rotating grain
are treated as nearly free, the magnetization of
the grain is the same for a stationary grain in a ``Barnett-
equivalent'' magnetic field directed along the grain angular 
velocity $\omega$
and having an amplitude 
$
H_{\rm BE}=\hbar\omega/(g\mu_{\rm B})
$, where $g$
is the electron gyromagnetic ratio $\approx 2$ and $\mu_{\rm B}$ is the Bohr magneton.
In these conditions the component of magnetic field perpendicular 
to $ \omega$ causes electron
spin resonance (see Athertson 1973).

LD00 called the process of paramagnetic relaxation within
a rotating body ``resonance relaxation'' as opposed to Davis-Greenstein
relaxation that disregards the spontaneous magnetization of a rotating body.
Solving the 
Bloch equations (Bloch 1946) LD00 obtained the  following expression
for
the imaginary part of the grain paramagnetic susceptibility (the
part responsible for dissipation and therefore alignment):
\begin{equation}
Im(\chi)=\chi_0 
\frac {\omega \tau_2}
{1+\gamma^2 g^2 \tau_1 \tau_2H_1^2}~~~,
\label{eq:chi''}
\end{equation}
where  $\gamma\equiv e/2 m_e c=8.8\times 10^6$~s$^{-1}$G$^{-1}$,
$\tau_1$ is the spin-lattice relaxation time, and $H_1$ is
interstellar magnetic field intensity.
Unlike the corresponding expression in Davis-Greenstein theory,  
eq.~(1) does not vanish for $\omega$ much larger
than the spin-spin relaxation time $\tau_2$. The saturation,
however, depends on the value of $H_1$
and $\tau_1$. The latter parameter
was calculated in LD00 using Raman scattering of phonons, but
the calculations are based on the so-called Waller theory, which
is known to overestimate $\tau_1$ considerably. Thus laboratory 
measurements of relaxation within isolated grains are required.

\begin{figure}
\includegraphics[width=1.0\textwidth]{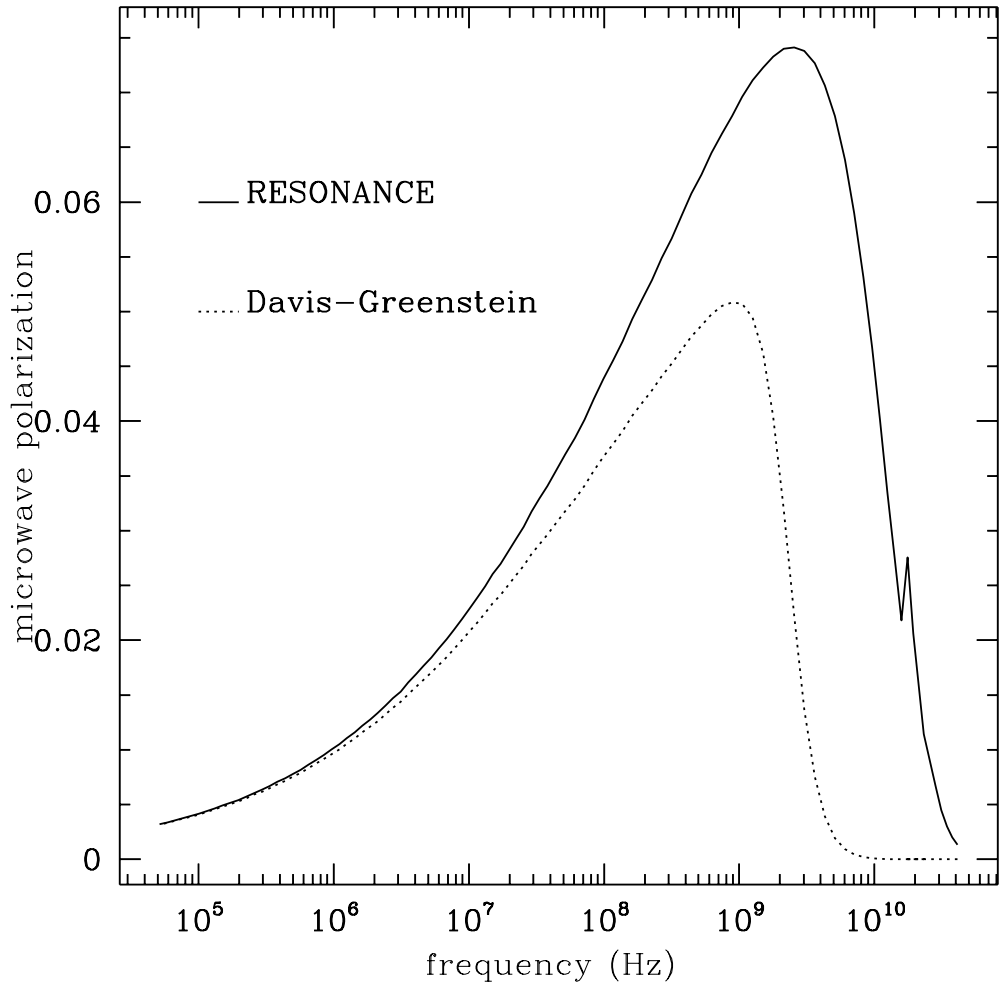}
\caption{Measure of grain alignment for both
	resonance relaxation and Davis-Greenstein relaxation for grains in
	the cold interstellar medium as a function of frequency (from LD00).
For resonance relaxation the saturation effects 
(see eq.\ (1)) are neglected, which means
that the upper curves correspond to the {\it maximal} values allowed by the 
paramagnetic mechanism.}
\end{figure}

Fig.~2 shows the predictions of the resonance relaxation mechanism
for cold interstellar gas
assuming that the spin-lattice relaxation is fast.  The discontinuity 
at $\sim 20$~GHz
is due to the assumption that smaller grains are planar, and larger
grains are spherical. The microwave emission will be polarized
in the plane perpendicular to magnetic field. 
The dipole rotational 
emission predicted in DL98a,b is sufficiently 
strong that polarization of a few percent
may interfere with efforts to measure the polarization of
the CMB.

Can we check the alignment of ultrasmall grains via infrared polarimetry? 
The answer to this
question is ``probably not''. Indeed, as discussed earlier,
infrared emission from ultrasmall grains, e.g. 12 $\mu$m emission,
takes place as grains absorb UV photons. These photons raise
grain temperature, randomizing grain axes in relation to
its angular momentum (see Lazarian \& Roberge 1997). Taking values
for Barnett relaxation from Lazarian \& Draine (1999a), we get
the randomization time of the $10^{-7}$~cm grain to be
 $2\times 10^{-6}$~s, which is less than grain cooling time. As the
result, the emanating infrared radiation will be polarized very marginally.
If, however, Barnett relaxation is suppressed, the randomization time 
is determined
by inelastic relaxation (Lazarian \& Efroimsky 1999) and is 
$\sim 0.1$~s, which would entail a partial polarization of
infrared emission.

\section{Polarization of Magneto-Dipole Emission}

While the spinning grain hypothesis got recognition in the 
community, the magnetic dipole emission model suggested by Draine
\& Lazarian (DL99) was left essentially unnoticed. 
This is unfortunate, as magnetic dipole emission provides a
possible alternative explanation to the Foreground X. Magnetic 
dipole emission is negligible at optical and infrared frequencies. 
However, when the
frequency of the oscillating magnetic field approaches the precession
frequency of electron spin in the field of its neighbors, i.e.
$10$~GHz, the magneto dipole emissivity becomes substantial.

How likely is that grains are strongly magnetic? Iron is the fifth
most abundant element by mass and it is well known that it resides in
dust grains (see Savage \& Sembach 1996). If $30\%$ of grain mass is
carbonaceous, Fe and Ni contribute approximately $30\%$ of the remaining
grain mass. Magnetic inclusions are widely discussed in grain
alignment literature (Jones \& Spitzer 1967, Mathis 1986, Martin 1995, 
Goodman \&  Whittet 1995). 
If a substantial part of this material is ferromagnetic
or ferrimagnetic, the magneto-dipole emission can be comparable to that
of spinning grains. Indeed, calculations in DL99 showed that less than
$5\%$ of interstellar Fe in the form of metallic grains or inclusions
is necessary to account for the Foreground X at 90~GHz, while magnetite,
i.e. Fe$_3$O$_4$, 
can account for a considerable part of the anomalous emissivity over
the whole range of frequencies from 10 to 90~GHz.

The mechanims of producing polarized
magneto-dipole emission is similar to that
producing polarization of electro-dipole  thermal emission 
emitted from aligned non-spherical grains (see Hildebrand 1988).
There are two
significant differences, however. First, strongly magnetic
grains can contain just a single magnetic domain. Further magnetization
along the axis of this domain is not possible and therefore the 
magnetic permeability of the grains gets anisotropic: $\mu=1$ along
the domain axis, and $\mu=\mu_{\bot}$ for a perpendicular direction.
Second, even if a grain contains tiny magnetic inclusions and can be
characterized by isotropic permeability, polarization that it produces
is orthogonal to the electrodipole radiation emanating through
electro-dipole vibrational emission. In case
of the electo-dipole emission, the longer grain axis defines the vector
of the electric field, while it defines the vector of the 
magnetic field in case
of magneto-dipole emission.

\begin{figure}[h!t!]
\includegraphics[width=1.0\textwidth]{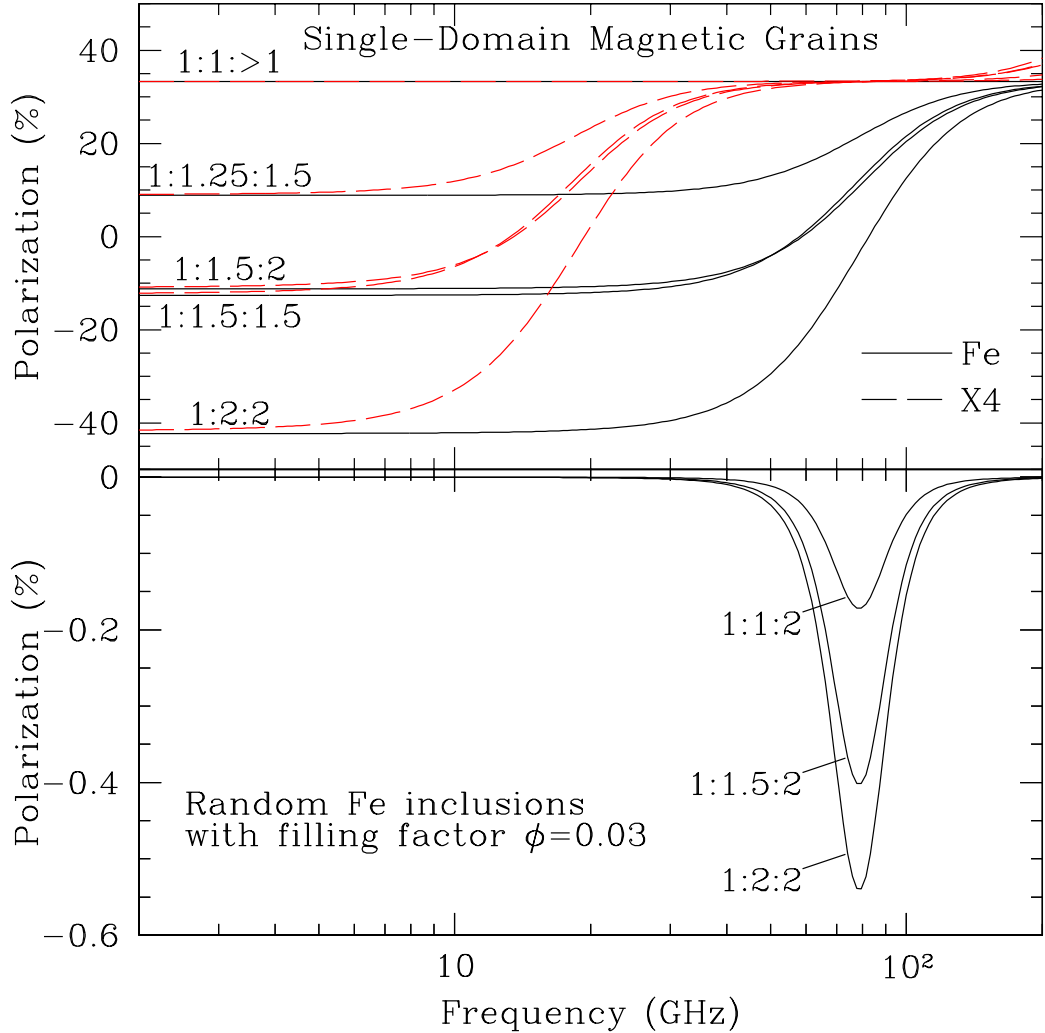}
\caption{ Polarization from magnetic grains (from DL99). Upper panel:
Polarization of thermal emission from perfectly aligned single
domain grains of metallic Fe (solid lines) or hypothetical magnetic
material that can account for the Foreground X (broken lines).
Lower panel: Polarization from perfectly aligned grains with
Fe inclusions (filling factor is 0.03). Grains are ellipsoidal and
the result are shown for various axial ratios.
}
\end{figure} 

The results of calculations for single domain iron particle (longer axis
coincides with the domain axis) and a grain with metallic Fe inclusions
are shown in Fig.~3. Grains are approximated by ellipsoids $a_1<a_2<a_3$
with ${\bf a_1}$ perfectly aligned 
parallel to the interstellar magnetic field ${\bf B}$. The polarization
is taken to be positive when the electric vector of emitted radiation
is perpendicular to ${\bf B}$; the latter is the case for electro-dipole
radiation of aligned grains. This is also true (see Fig.~3) for high
frequency radiation from single dipole grains. It is easy to see
why this happens. For high frequencies $|\mu_{\bot}-1|^2\ll 1$
and grain shape factors are unimportant. The only important thing is
that the magnetic fluctuations happen perpendicular to ${\bf a_1}$. 
With ${\bf a_1}$ parallel to ${\bf B}$, the electric fluctuations
tend to be perpendicular to ${\bf B}$ which explains the polarization
of single domain grain being positive. For lower frequencies magnetic
fluctuations tend to happen parallel to the intermediate size axis 
${\bf a_2}$. As the grain rotates about ${\bf a_1}\|{\bf B}$,
the intensity in a given direction reaches maximum when an observer
sees the ${\bf a_1} {\bf a_2}$ grain cross section. Applying earlier
arguments it is easy to see that magnetic fluctuations are parallel
to ${\bf a_2}$ and therefore for sufficiently large $a_2/a_1$ ratio
the polarization is negative. {\it 
The variation of the polarization direction with
frequency presents the characteristic signature of magneto-dipole emission
from aligned single-dipole grains and it can be used to separate this
component from the CMB signal}. Note that the degree of polarization is
large, and such grains may substantially interfere with the attempts
of CMB polarimetry. Even if the
intensity of magneto-dipole emission is subdominant
to that from rotating grains, it can still be quite important in
terms of polarization.
A relatively weak polarization response is expected for grains with
magnetic inclusions (see Fig.~3). The resulting emission is negative
as magnetic fluctuations are stronger along longer grain axes, while
the short axis is aligned with ${\bf B}$.

\section{Polarized Emission from the Galaxy}
Attempts to determine the statistics of intensity and polarization
fluctuations of cosmic microwave background (CMB) renewed
interest to the fluctuations of Galactic foreground radiation
(see Tegmark et al.~2000).
Angular power spectra of intensity of synchrotron emission 
and synchrotron polarization
(see papers in de Oliveira-Costa \& Tegmark 1999)
as well as starlight polarization 
(Fosalba et al. 2002; henceforth FLPT)
have been measured. 
Those measurements revealed a range of power-laws.\footnote{
     It is customary for
     CMB studies to expand
     the foreground intensity  over spherical harmonics $Y_{lm}$,
        $I(\theta, \phi)= \sum_{l,m}a_{lm}Y_{lm}(\theta,\phi)$, 
     and write the spectrum in terms of  
        $C_l\equiv \sum_{m=-l}^{m=l} |a_{lm}|^2/(2l+1)$.
    The measurements indicate that 
    angular power spectrum ($C_l$) of the Galactic emission
    follows power law ($C_l\propto l^{-\alpha}$)
   (see FLPT).
    The multipole moment $l$ is related to the angular scale 
    $\theta$ on the sky as
    $l\sim \pi/\theta^{radian}$
    (or $l\sim 180^{\circ}/\theta^{\circ}$).
   When the angular size of the observed sky patch
   ($\Delta \theta \times \Delta \theta$ in radian)
   is small, 
    $C_l$ is approximately the `energy' spectrum of fluctuations
   (Bond \& Efstathiou 1987).
   It is this power spectrum expressed in terms of wavenumber
   $k ~(\sim l\, \Delta \theta/\pi)$ 
   that is usually dealt with in studies of astrophysical
   turbulence (e.g. Stanimirovic \& Lazarian 2001).   \label{footnote10}
    }

Interstellar medium is turbulent and Kolmogorov-type spectra 
($E_{3D}(k)\sim k^{-11/3}$, $k$=wave number)
were reported on the scales from several AU to several kpc
(see Armstrong, Rickett, \& Spangler 1995; Lazarian \& Pogosyan 2000; 
Stanimirovic
\& Lazarian 2001). Therefore it is natural to think of the 
turbulence as the origin of the fluctuations of the diffuse foreground
radiation. Interstellar medium is magnetized with magnetic
field making turbulence anisotropic. 
Recent developments on MHD turbulence can be found in
Goldreich \& Sridhar (1995), Lithwick \&
Goldreich (2001), and Cho \& Lazarian (2002a).
See also a review Cho, Lazarian \& Vishniac (2002).

The polarization of starlight and of the Far-Infrared
Radiation (FIR) from aligned dust grains is affected by the ambient
magnetic fields. 
Therefore, the structure of MHD turbulence is closely related to
the polarization observations.
Assuming that dust grains are always aligned with
their longer axes perpendicular to magnetic field (see the review
Lazarian 2000), one gets the 2D distribution of the magnetic field
directions in the sky. Note that the alignment is a highly non-linear
process in terms of the magnetic field and therefore the magnetic
field strength is not available\footnote{%
The exception to this may be the alignment of small grains which can
be revealed by microwave and UV polarimetry (Lazarian 2000).
}. 

The statistics of starlight polarization (see FLPT)
is rather rich for the Galactic plane and it allows to establish the
spectrum\footnote{%
Earlier papers dealt with much poorer samples (see Kaplan \& Pickelner
1970) and they did not reveal power-law spectra.
} \( C_l\sim l^{-1.5} \), where \( l \) is the multipole moment (see footnote
\ref{footnote10}).
For uniformly sampled turbulence it follows from Lazarian \& Shutenkov
(1990) that \( C_l\sim l^{\alpha } \) for \( l<l_{0} \) and \( l^{-1} \)
for \( l>l_{0} \), where \( 180^{\circ}/l_{0} \) is the critical angular
size of fluctuations which is proportional to the ratio of the injection
energy scale to the size of the turbulent system along the line of
sight. For Kolmogorov turbulence \( \alpha =-11/3 \). 

However, the real observations do not uniformly sample turbulence.
Many more close stars are present compared to the distant ones. Thus
the intermediate slops are expected. Indeed, Cho \& Lazarian (2002b)
showed through direct simulations that the slope obtained in FLPT (2002)
is compatible with the underlying Kolmogorov turbulence.
At the moment FIR polarimetry does not provide maps that are really
suitable to study turbulence statistics. This should change soon when
polarimetry becomes possible using the airborne SOFIA observatory.
A better understanding of grain alignment (see Lazarian 2000) is required
to interpret the molecular cloud magnetic data where some of the dust
is known not to be aligned (see Lazarian, Goodman, \& Myers 1997 and
references therein). 

Another way to get magnetic field statistics is to use synchrotron
emission. Both polarization and intensity data can be used. The angular
correlation of polarization data (Baccigalupi et~al. 2001) shows
the power-law spectrum \( l^{-1.8} \) and we believe that the interpretation
of it is similar to that of starlight polarization. Indeed, Faraday
depolarization limits the depth of the sampled region. The intensity
fluctuations were studied in Lazarian \& Shutenkov (1990) with rather
poor initial data and the results were inconclusive. Cho \& Lazarian
(2002b) interpreted the fluctuations of synchrotron emissivity (Giardino
et~al. 2001, 2002) in terms of turbulence with Kolmogorov spectrum.

\section{Summary}

The principal points discussed above are as follows:
\begin{itemize}

\item Grain alignment theory has at last reached its mature state
when predictions for classical ($a>3\times 10^7$~cm) grains
are possible. In most cases grain alignment happens
in respect to magnetic field, i.e. reveal magnetic field direction,
even if the alignment mechanism is not
magnetic. However, depending on the mechanism the alignment may
happen with grain longer axes parallel or perpendicular to magnetic
field.  

\item ``Resonance paramagetic relaxation'' is the process that can enable
alignment and therefore polarization from ultrasmall grains. Although
the details of the process still require laboratory testing, it looks
that the polarization is marginal beyond 40~GHz. This process requires
further work. 

\item Magneto-dipole emission may be
substantially polarized in the range 10-90~GHz.
Thus it can be important in terms of polarization even if it is
subdominant in terms of total emissivity. 

\item Statitics of the galactic foreground polarization may be interpreted
in terms of underlying MHD turbulence.

\end{itemize}

Support by NSF grant AST-0125544 is acknowledged.

\end{document}